\documentstyle[12pt]{article}
\newcommand{\bm}{\bibitem}

\begin{document}
\thispagestyle{empty}
\headsep 1.5 cm
\textheight 22.0  cm
\normalbaselineskip = 24 true pt
\normalbaselines
\rightline{\large\sf SINP-TNP/95-23}

\rightline{\large\sf November 1995}

\begin{center}
{\Large\bf
{ Symmetry breaking for rho meson in neutron matter}}
\vskip 0.2 cm
{\sf Abhee K. Dutt-Mazumder \footnote{E-mail:
abhee@tnp.saha.ernet.in},  Anirban Kundu, \\
Triptesh De and Binayak Dutta-Roy }\\
Saha Institute of Nuclear Physics, 1/AF  Bidhan Nagar\\
Calcutta - 700 064, India\\
\end{center}
\begin{abstract}

Qualitative changes in the collective excitation spectra of the
$\rho$-meson triplet in neutron matter is studied, with particular
emphasis on the breaking of the discrete $p\leftrightarrow n$ symmetry.
The appearance of additional branches in the dispersion
characteristics, the mass splitting among the charge states,
the splitting between
longitudinal and transverse modes of the $\rho^{\pm}$ mesons and the
appearance of `island' modes (or loops) in the time-like region are
some of the features that are exposed.
\vskip 0.5 cm
\noindent{{PACS numbers: 12.38.Mh, 11.30.Rd, 12.40Vv, 21.65.+f}}\\
{\it{Keywords:}} $\rho$ meson, neutron matter,
collective oscillation\\
\end{abstract}

\newpage

In the recent past there have been several attempts both in the arena
of theory and experiment to unravel the behaviour of different
particles inside dense nuclear medium,
particularly at densities 2-3 times higher than the normal
nuclear matter density. Such studies are of cardinal importance both in
the context of heavy ion collision and nuclear astrophysics. These apart,
understanding of their properties inside hot and dense nuclear matter
would also pave the way for extracting valuable informations from the various
signals produced in the high energy heavy ion collisions needed to determine
the equation of state of the system. Here we study the
in-medium behaviour of the $\rho$ meson in particular, as its significant role
in heavy ion collisions in connection with the dilepton spectroscopy has been
pointed out by several authors \cite{dls,as}. Of specific interest here is to
investigate the qualitative changes in the dispersion characteristics
when the
isospin symmetry is broken. Here
the Lagrangian respects the symmetry but it is
broken by the isospin asymetric ground state or the `vacuum'.

Earlier, the in-medium mass shift of the $\rho$ meson was investigated by
several authors, both at zero and finite temperature within the framework of
different theoretical models \cite{as,je,sh}. These studies
were mostly confined to isospin symmetric nuclear matter.
This letter, in contrast, as mentioned already,
attempts to probe  the properties of $\rho$ meson propagator in
nuclear matter where the so-called `vacuum' or the ground state does not
respect isospin symmetry, {\em i.e.}
the densities of the neutron and proton
are not equal. In order to bring out the effects in clear relief
we consider the extreme limit of asymmetry viz., neutron matter.
Our essential focus is on the collective oscillations
of neutron matter set by the propagating $\rho$ meson, popularly known as
particle modes \cite{ch}. Among the
features qualitatively different from what are observed in symmetric nuclear
matter, most remarkable is the splitting of longitudinal (L)
and transverse (T) modes in the collective excitation spectra
for $\rho^{\pm}$ meson  even in the static limit
and degeneracy of the same for $\rho^0$.
Also noteworthy is the appearance
of the little `islands' in the dispersion curves for $\rho^{\pm}$ mesons.
These branches represent two strongly interacting modes that
survive only upto a finite value of momentum having a minimum at
somewhat lower value of $|\vec q|$ (Fig. 1). Also significant is the
appearance of additional branches in the excitation spectra
usually absent for symmetric nuclear matter.
\input psbox.tex
\psbox{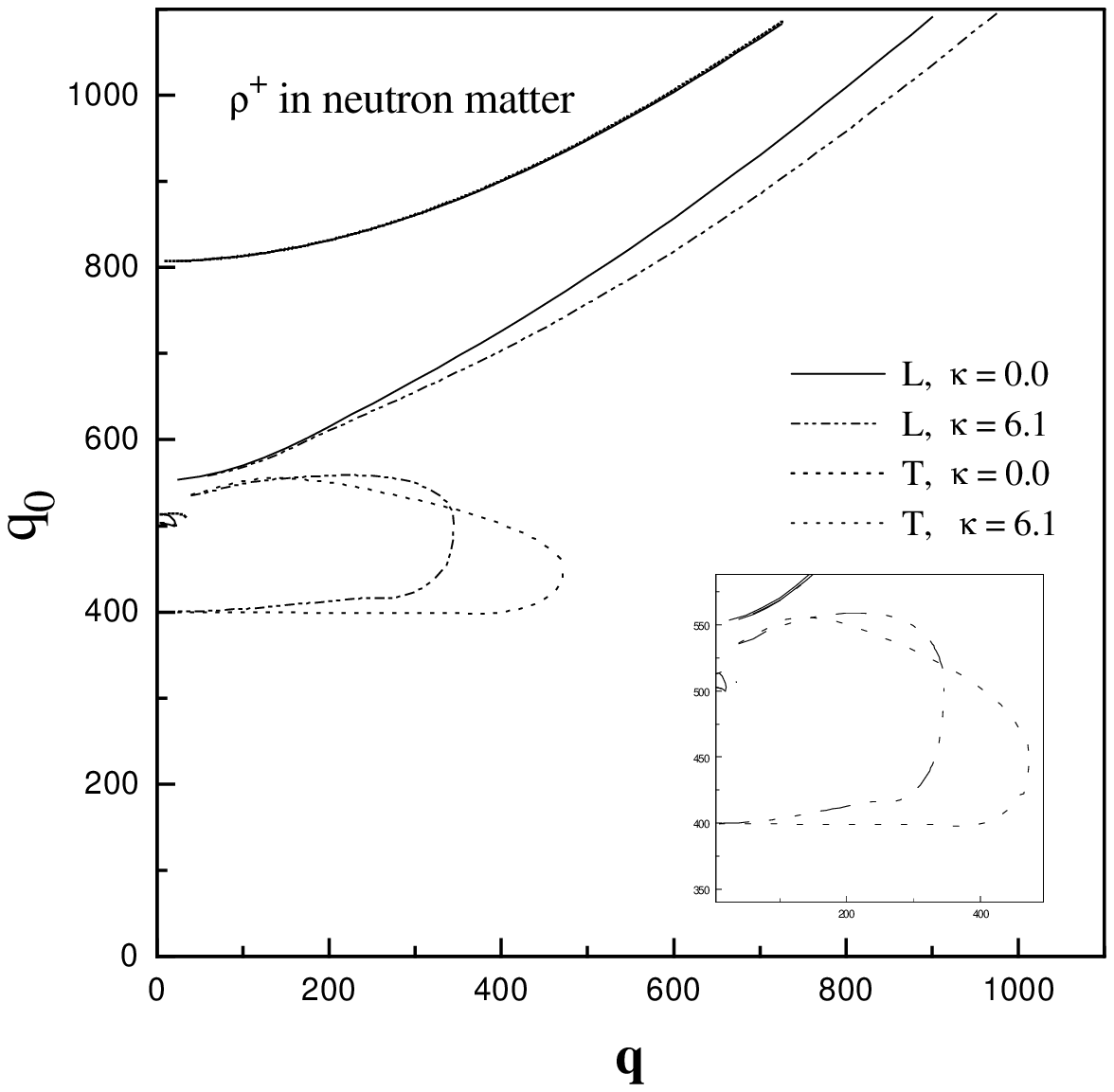}
The Lagrangian describing $\rho$-nucleon interaction may be written as
\begin{equation}
{\cal L}_{int} = g_v^j [{\bar{N}} \gamma _\mu \tau^j
     N - \frac{\kappa}{2M}{\bar{N}}
	 \sigma_{\mu\nu}\tau^j N\partial ^\nu]
\rho^\mu_j
\end{equation}
in presence of a tensor or magnetic interaction apart from the usual vector
interaction of the original Walecka model (QHD-I). The index $j$ runs
from 1 to 3, and we use $\rho^{\pm}=(\rho_1\mp i\rho_2)/\sqrt 2 $ and
$\rho_0=\rho_3$, $\tau ^j$ are the usual $2\times 2$ Pauli isospin
matrices, and
N denotes the two-component isospinor for the
nucleon. The coupling
constant $g_v$ and $\kappa$ may be estimated from the Vector Meson Dominance
(VMD)
of nucleon form factors or from the fitting of the nucleon-nucleon interaction
data as done by the Bonn group \cite{bonn}. From the estimates it is clear
that in case
of the
vector-isovector $\rho$ meson, the tensor interaction is substantial
\cite{song}. While VMD fits suggest the value $\kappa=3$, we consider the
larger estimates of the Bonn group again to emphasize the effects:
\begin{equation}
\frac{g_v^2}{4\pi} = 0.92 ~~~~~\kappa = 6.1
\end{equation}

As the detailed treatment of the underlying formalism is available elsewhere,
here, for brevity, we sketch only a brief outline of the same and
present the final results of our calculations. The essential idea here is to
determine the in-medium dressing of the $\rho$ meson propagator due to the
nucleon-hole and nucleon-antinucleon excitations (the former being the
effect of the Fermi sea, the latter driven by the Dirac vacuum).
Different $\Pi$-functions or vacuum polarizations are therefore calculated
to study the in-medium effects on the $\rho$ meson propagation inside the
nuclear matter. We concentrate
only on the effect of the Fermi sea which we refer as `dense part', the
contribution of the Dirac sea on the other hand is not included in the present
analysis. The form of the nucleon propagator in nuclear matter may be expressed
as
\begin{equation}
G(k) = G_F(k)  + G_D(k)
\end{equation}
where
\begin{equation}
G_F(k) =
(k\!\!\!/ +M^\ast)[\frac{1}{k^2-M^{\ast2}+i\epsilon}]
\end{equation}
and
\begin{equation}
G_D(k) = (k\!\!\!/ + M^\ast) [\frac{i\pi}{E^{\ast}(k)}
\nonumber\\ \delta (k_0-E^\ast (k))\theta(k_F-|\vec k |)]
\end{equation}
with $M^\ast$ denoting the effective mass of the nucleon in the
medium and $E^\ast (\vert \vec k \vert) = \sqrt {\vert \vec k \vert^2 +
M^{\ast 2}}$.
The first term in $G(k)$, namely $G_F(k)$,
is the same as the free propagator of a spin $\frac{1}{2}$
fermion, except for the fact that the effective mass of the nucleon
is to be used, while the second part, $G_D(k)$,
involving $\theta(k_F-|\vec k|)$,
arises from Pauli blocking, describes the modifications
of the same in the nuclear matter at zero temperature \cite{se}, as it
deletes the on mass-shell propagation of the nucleon in nuclear
matter with momenta below the Fermi momentum. Finite temperature effects
may be incorporated by replacing the Heaviside $\theta$-function by
the Fermi distribution, but the main features to be exposed in
this paper are not modified significantly.
In general, the self-energy
or the second order polarization function may be written as
\begin{equation}
\Pi_{\mu\nu}^{lm}=\frac{-i}{(2\pi)^4}
\int {d^4}k~{\rm Tr}[i\Gamma^l_\mu iG(k+q)
{i\bar\Gamma^m_\nu} iG(k)]
\end{equation}
where $\Gamma$ represents appropriate vertex factors and $l$,$m$ are the
isospin indices. Just like the nucleon
propagator, the polarization function can also be expressed as a sum
of two parts, one coming from the dense part contribution and the other being
the contribution of the Dirac sea. Here we consider only the first.
\begin{equation}
\Pi_{\mu\nu}(q)= \Pi_{\mu\nu}^F(q) + \Pi_{\mu\nu}^D(q)
\end{equation}

The real part of the
density dependent polarization tensor for $\rho^+$ propagation
is given by
\begin{eqnarray}
\Pi^D_{\mu\nu}
&=&\frac{g_v^2}{(2\pi)^3}\Big [\int_0^{k_F^{(p)}}\!\!\!\frac{d^4k}
{E_p^{\ast}(k)}
\delta (k_0-E_p^\ast (k))\frac{{\it {T}}_{\mu\nu}(k-q,k)}
{(k-q)^2-M_n^{\ast 2}}\nonumber\\
&{ }&+ \Big (k\rightarrow k+q, p\rightarrow n\Big)\Big ]
\end{eqnarray}
where the superscript on $k_F$ and the subscript on $E^{\ast}$ and
$M^{\ast}$ denote the corresponding quantities for protons and
neutrons.
$T_{\mu\nu}$ is the relevant trace taken over the fermion loop, the
arguments being the fermionic momenta. Results for $\rho^-$ are obtained
by making the replacement $q_0\rightarrow -q_0$ and $|\vec q|\rightarrow
-|\vec q | $. The angle integration of eq. (8) is performed analytically
whereas the momentum integration is computed numerically.

The collective modes arising out of the meson propagation in neutron
matter is characterized by the zeros of the dielectric functions,
$\epsilon(q_0,|\vec q|)$,
which is calculated from the self energy \cite{ch}. Independent
components of $\Pi_{\mu\nu}$ are combined appropriately to give longtudinal
and transverse modes of the dielectric functions. Therefore the
eigenconditions for collective oscillations are given by
\begin{equation}
\epsilon_T^2(q) \epsilon_L(q) = 0
\end{equation}
Dispersion characteristics corresponding to the
longitudinal and transverse modes are obtained by  solving this equation
for $q_0$ and $|\vec q|$ numerically.

It is evident from eq. (8) that the density dependent part of the
polarization involves two integrations having two different limit
of integration corresponding to neutron and proton fermi momenta.
In symmetric nuclear matter the radii
of two fermi spheres (proton and neutron) are the same, in other words
discrete isospin symmetry $p\leftrightarrow n$ remains exact in this limit.
Hence the effective masses of the neutron and proton are also equal. This
ensures the conservation of current for all the isospin multiplets
in symmetric nuclear matter, which in turn implies that to every longitudinal
mode, there exists a corresponding transverse mode which coincide in
the static limit ($|\vec q|=0$) \cite{abhee}.

This feature, however, is not totally present
in nuclear matter having different
number of neutron and proton, that is to say that the
symmetry is broken resulting in the splitting of the longitudinal
and transverse modes even in the static limit. This situation is
noteworthy as in this case even though the Lagrangian is invariant
under rotation in isospin space, the `vacuum' is not, and
hence particle
spectrum also breaks the isospin symmetry.
However, for $\rho^0$ we
have n-n and p-p loops, even for asymmetric
nuclear matter having two fermi spheres of
different sizes (one may be zero)
and, therefore, in the loop integral effective masses involved are either
$M_p^\ast$ or $M_n^\ast$ making it possible to
combine the integrations so as
to express it in a `gauge invariant' form.
It may be shown that in this case the requirement of current conservation
is fulfilled {\em i.e. }, $q^\mu \Pi_{\mu\nu}=\Pi_{\mu\nu}q^\nu = 0 $,
while in case of $\rho^\pm$ this criteria is satisfied only if $k_F^{(n)} =
k_F^{(p)}$.
Hence, for $\rho^0$ propagation, no matter whether the
ground state of the nuclear system maintains this symmetry or not, in the
static limit the branches corresponding
to the longitudinal and transverse mode
always coincide ${\it i.e.}~ \rho^0_{trans}$ and $\rho^0_{long}$ have
same value for the
mass even in neutron matter. We refer our readers to Fig. 2.
However, this mass changes with the value of
$\kappa$ and density.
\input psbox.tex
\psbox{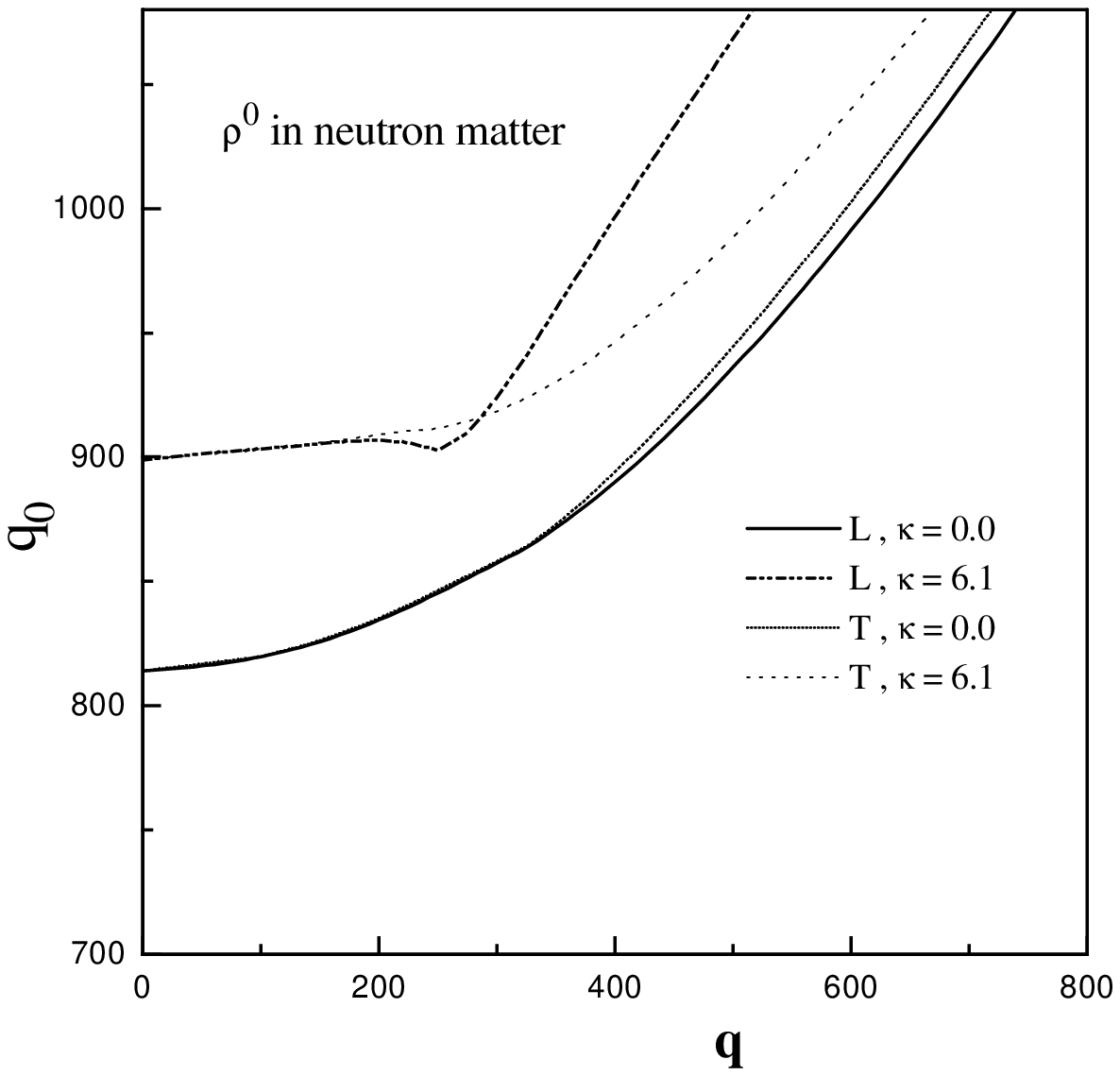}
For the virtual proton propagator in neutron matter only the first part,
viz. $G_F$
survives whereas the second part, $G_D$, vanishes as $k_F^{(p)}\rightarrow
0$. Of course
for the neutron propagator both the terms are nonvanishing.
As in the present case we have only the neutron fermi sphere,
$\rho^+$ and $\rho^-$ propagators will be modified differently. For
the former the modification is due to proton-neutron hole, while the
latter is being dressed by the neutron-antiproton excitation. In case
of $\rho^0$, however, proton-antiproton and neutron-neutron hole will
contribute as pointed out earlier. In the $\rho^0$ spectra, it is observed
that for $\kappa=6.1$, though
the longitudinal and transverse modes cross each
other, mode conversion  is not possible as the nature of
polarization is different. It should also be noted that
the degeneracy of these
two modes in the static limit follows from our earlier remark.

Next we focus on the $\rho^-$ propagation. The dispersion characteristics
in the region of stability , {\em i.e.},
 for $q_0\leq (M_p+E_F^n)$ (beyond which
the collective modes decay via particle-hole excitations),
have been depicted in Fig. 3.
\input psbox.tex
\psbox{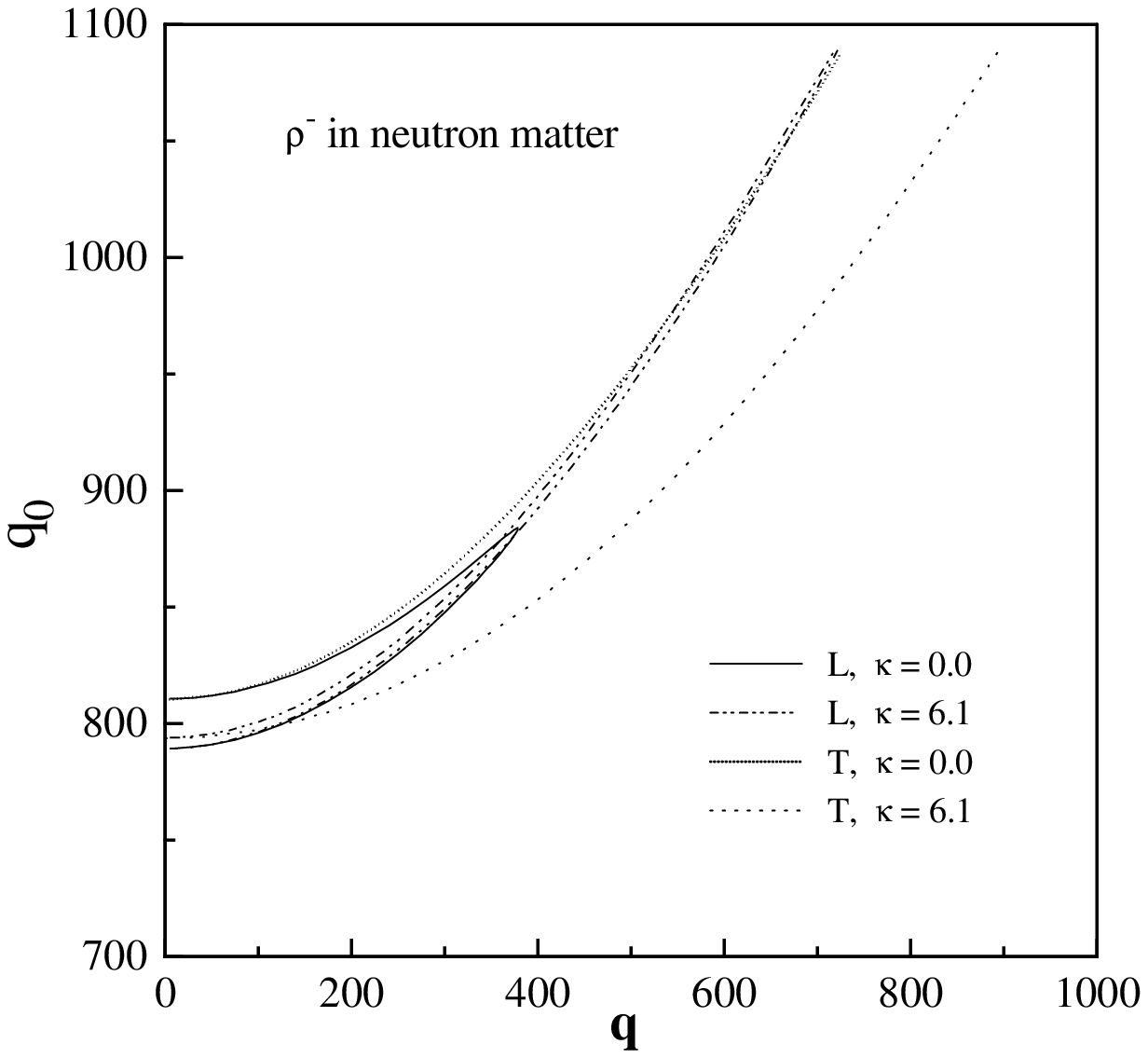}
Unlike the case of $\rho^0$, here the
longitudinal and transverse modes are not necessarily degenerate in
the static limit. Even more interesting is the case of $\rho^+$
propagation. Depending on the strength of magnetic interaction one can have
at most four L, and three T modes in the region of stable collective
oscillation (from the charge symmetry one can say that the situation is
similar to that of $\rho^-$ propagation in proton matter). This is evident
from Fig. 2 where one of the longitudinal modes is unpaired. It is worth
noting that the separation between the paired modes
increases with the introduction
of $\kappa$. The appearance of the so-called `island'
modes in the particle branch, which to the best of our knowledge has
not been highlighted earlier (though such a situation has been encountered
in the spacelike region \cite{zero}), is another feature that we want to
emphasize.  For $\rho^+$, the T and L branches
corresponding to the `island' modes are coincident in the static limit.

One of the interesting property of these  modes
is their presence even in the limit $\kappa\rightarrow 0$ both
for $\rho^+$ and
$\rho^-$ propagation. However the islands
are  absent in $\rho^0$ spectra. Such modes are not
observed in case of symmetric nuclear matter, hence these are to be understood
as the effect of asymmetry. In case of $\rho^-$ in neutron matter again
these `islands' are not found for the
transverse while for the longitudinal modes
all of them form loops. $\rho^+$ on the other hand shows loops both for
longitudinal and transverse polarizations. The loops grow in size with $\kappa$
and with the density of nuclear matter.
It is appropriate to note the shape of the `islands';
for $\rho^-$ these are narrow and sharp at the end while for $\rho^+$ they
are rounded.
Coupling of the modes or the appearance of these
loops indicate the onset of instability beyond a
particular value of momenta ($|\vec q|_{max}$). In fact beyond this limit
the roots of $q_0$ become complex.   For completeness we have also shown the
variation of $|\vec q|_{max}$ as a function of density. We find that
$|\vec q|_{max}$ approaches a saturation value with increasing density, which
is shown in Fig. 4.
\input psbox.tex
\psbox{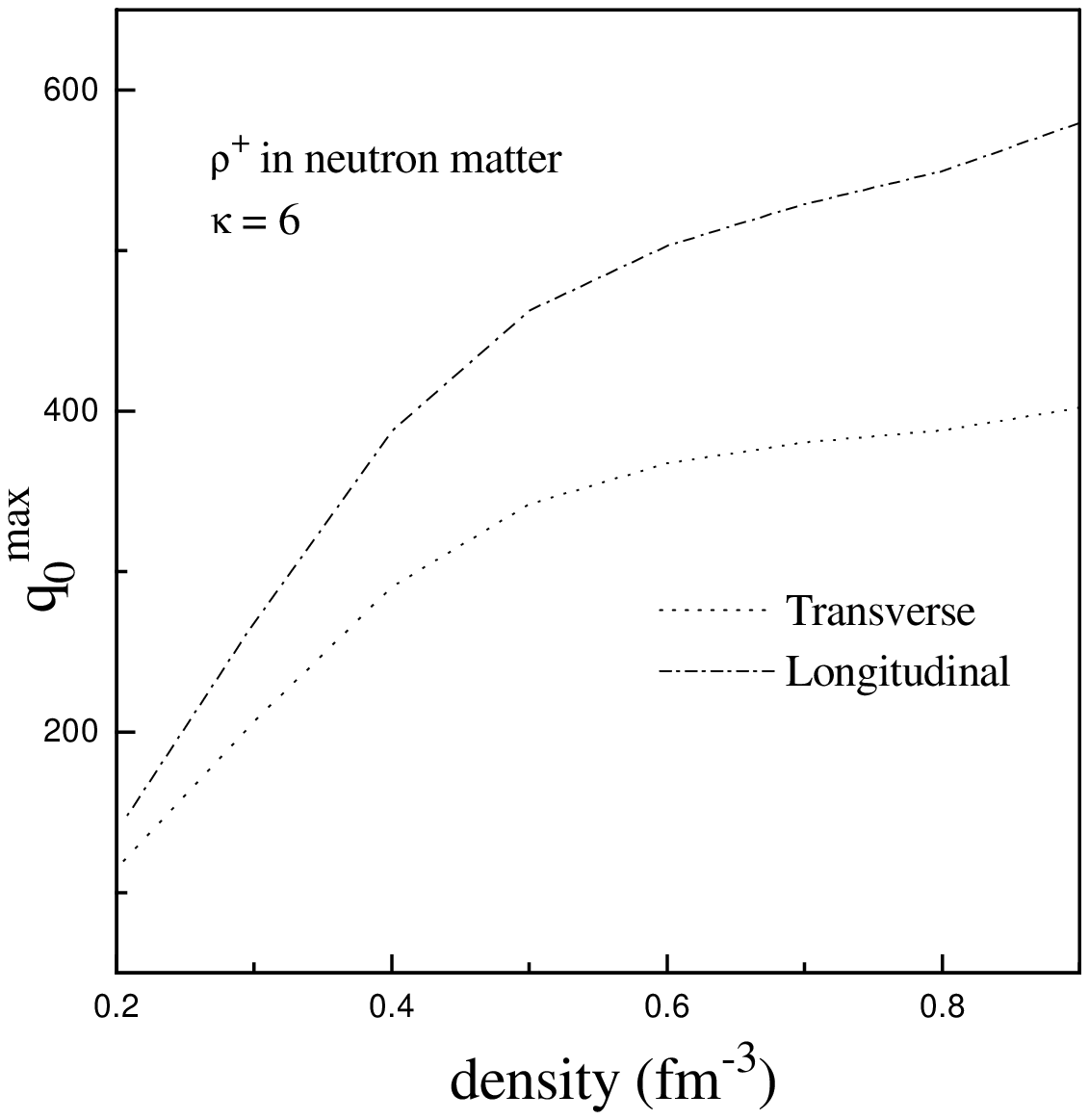}
To conclude and summarize we note that the $\rho$ meson propagation in neutron
matter is qualitatively different from what is obtained
in symmetric nuclear matter.
However for $\rho^0$, results do not alter much as expected from the fact that
$\rho^0$ is blind to isospin asymmetry. While on the other hand $\rho^+$ and
$\rho^-$ propagators are modified differently by two distinct mechanism
viz., proton-hole and neutron-antiproton pair creation respectively, and
therefore corresponding dispersion curves also reflect this fact.
It is observed that when the $p\leftrightarrow n$
symmetry is broken, additional `island' modes appear, together
with the $\rho^0$, $\rho^\pm$ mass splitting and the removal of
the degeneracy of L and T branches (even in the static limit).
Of course for more reliable quantitative estimates,
the contribution of the free part {\em i.e.}, the modifications due
to the Dirac sea and the effect of mixing have to be incorporated.
However, as we focus here only on the
qualitative changes in the dispersion characteristics, the effect of Dirac
sea is not expected to cause much difference as far as the nature of the
collective oscillations are concerned. It can be easily shown that for the
free part, the contribution to the self-energy is expressible as
$\Pi^F_{\mu\nu}=Q_{\mu\nu}\Pi$, where $Q_{\mu\nu}=-g_{\mu\nu}
+q_{\mu}q_{\nu}/q^2$. This evidently does not cause any splitting between
the L and T branches in the static limit. Apart from that, and only
for the diagram where both the $\rho\bar N N$ couplings are of vectorial
nature, there is an extra term which is proportional to $g_{\mu\nu}
\delta m$, where $\delta m$ is the mass difference between the two
fermions in the loop, and is nonvanishing only for asymmetric matter.
Studies are in  progress to investigate
dispersion characteristics in a varied range of
$p\leftrightarrow n$ asymmetry.
Detailed results will be presented elsewhere.

\newpage

\centerline{\bf Figure Captions}
\begin{enumerate}

\item Dispersion characteristic for $\rho^+$ meson (particle modes).
The L and T modes for $\kappa=0.0$ are nearly coincident. Both $q_0$
and {\bf q} are in MeV. The `island' modes are shown in detail.

\item Dispersion characteristics for $\rho^0$ meson (particle modes).
For $\kappa = 6.1$, the L and T modes cross.

\item Dispersion characteristic for $\rho^-$ meson  (particle modes).
The L mode for $\kappa=0$ form `island', and for $\kappa=6.1$ form
`island' just outside the stable oscillation region.

\item Change of $q_0^{max}$ (for `island' modes and $\kappa=6.1$) with
density.

\end{enumerate}

\end{document}